\documentclass[preprint,aps,eqsecnum,showpacs]{revtex4}
\usepackage{graphicx}
\usepackage{latexsym}
\begin{document}
\title{Brane cosmology, Weyl fluid, and density perturbations}
\author{Supratik Pal \footnote{Electronic address: {\em{supratik\_v@isical.ac.in}}}}
${}^{}$
\affiliation{Physics and Applied Mathematics Unit \\ 
Indian Statistical Institute \\
203 B.T.Road, Kolkata 700 108, India}
\newcommand{\be}{\begin{equation}}
\newcommand{\ee}{\end{equation}}
\newcommand{\bea}{\begin{eqnarray}}
\newcommand{\eea}{\end{eqnarray}}
\newcommand{\bml}{\begin{subequations}}
\newcommand{\eml}{\end{subequations}}
\newcommand{\bfig}{\begin{figure}}
\newcommand{\efig}{\end{figure}}

\vspace{.5in}

\begin{abstract}

We develop a technique to study relativistic perturbations in the generalised brane
cosmological scenario, which is a generalisation of the multi-fluid cosmological
perturbations to brane cosmology. The novelty of the technique lies 
in the inclusion of a radiative bulk which is responsible for
bulk-brane energy exchange, and in turn, modifies the standard perturbative analysis
to a great extent.
The analysis involves a geometric fluid -- called the Weyl fluid --
whose nature and role have been studied extensively both for the empty bulk and
the radiative bulk scenario.
Subsequently, we find that this Weyl fluid can be a possible geometric candidate for dark
matter in this generalised brane cosmological framework.

\end{abstract}

\pacs{04.50.+h, 98.80.-k, 98.80.Cq}

\maketitle

\section {Introduction}

During the last few years braneworld gravity has emerged as a 
more general theory of gravity, mainly due to the
possibility of explaining the gravitational phenomena observed in the four dimensional
universe from a  broader perspective \cite{eee, maartrev}. Subsequent developments
of the theory in the cosmological sector \cite{langrev}
came as an  inevitable outcome since the challenges
any theory of cosmology, be it a theory based on General Relativity
or any other phenomenologically motivated theory, faces  in explaining predictions
from the highly accurate  observational data \cite{cmb, dm, sdss}.
In spite of great complications involved, 
the cosmological aspects of this scenario
did show some promising features. To mention a few, brane cosmology naturally gives
rise to singularity-free bouncing and cyclic universes \cite{bounce}.
Also, in this theory, the universe does not need any special initial condition
for the inflation to start so that the isotropy is built in the theory
\cite{iso}. Even the possibility of inflation without any 4D  inflaton  field
is in vogue \cite{infl}.
Brane cosmology thus results in interesting physics which needs further investigations.

In this scenario, the bulk spacetime is either AdS$_5$  \cite{ads5} or 
a generalised version of it. The generalised  global structure  depends upon 
whether the bulk has only a cosmological constant 
or there is  any non-standard model fields minimally or non-minimally
coupled to gravity or to brane matter.
When the bulk is empty consisting only of a cosmological constant, the bulk metric in which an 
FRW brane can be consistently embedded, is given by a 5-dimensional Schwarzschild-Anti de Sitter (Sch-AdS$_{5}$) or a Reissner-N\"{o}rdstrom Anti de Sitter (RNAdS$_{5}$)
black hole  \cite{maartrev,langrev, scads5, genglob, bounce} . A subsequent generalisation of this scenario can be obtained when the bulk is not necessarily empty but it consists of a radiative field,
resulting in a Vaidya-Anti de Sitter (VAdS$_5$)  black hole for the bulk metric
\cite{maartbulk, langrev, lang1, chamb, lang2, vads5gen, asymgen, nonradial, radbh5, sustr, sucol}.
A `black hole in the bulk' scenario provides us with a novel way of visualising cosmological phenomena on the 4D universe.
In this scenario, the brane is moving in the bulk, with its radial trajectory being
identified with the scale factor of the 4D world, so that the expansion of the
universe is a realisation of the radial trajectory of the brane in the bulk.

The most notable contribution from bulk
geometry on the brane is, perhaps, an additional term in the Friedmann equations,
which arise from the projection of the bulk Weyl tensor onto the brane.
The precise role of this term, compatible to FRW background on the brane, 
is to supply a geometric perfect fluid whose nature is governed by the contents of the bulk
(in turn, bulk geometry)
we choose. For an empty bulk, it is radiation-like and is called the {\em dark
radiation}. 
There is extensive study in the literature either
by setting it to zero for practical purpose or by attributing a very small
value to it, constrained by Nucleosynthesis data ($< 3 \%$ of total radiation
energy density of the universe) \cite{maartrev, maartprog}. Examples include
metric-based perturbations \cite{pertmet}, density perturbations on
large scales neglecting dark radiation \cite{pertden}, or including its effects \cite{pertden2},
  curvature perturbations  \cite{pertcur} and the
Sachs-Wolfe effect \cite{brsachs}, vector perturbations
\cite{pertvec}, tensor perturbations \cite{pertten} and CMB
anisotropies \cite{pertcmb}. In all the cases, the effect has been
found to be modify the standard analysis very little, as expected
from its radiation-like behaviour.

On contrary, when the bulk is not necessarily empty,
the nature of this entity is no longer radiation-like, rather
it depends upon the contents in the bulk,
which is reflected by the VAdS$_5$ bulk geometry
\cite{maartbulk, langrev, lang1, chamb, lang2, vads5gen, asymgen, nonradial, radbh5, sustr, sucol,
ansatz, excos4}. 
It is thus important to determine the nature as well as
the role  of this entity, called in general the  {\em Weyl fluid},
 in the cosmological dynamics and perturbations,
and find if this scenario has some advantages over others.
A recent work \cite{sustr} has shown its significance as a possible dark matter candidate 
by Newtonian analysis of perturbations, followed by some confrontation with observations
\cite{obs}. However, as in GR, the Newtonian analysis of gravitational instability
is limited in the sense that it
fails to account for the perturbations on scales larger than the Hubble radius. One needs
relativistic analysis  valid at super-Hubble scales as well. 
Further, in order to test braneworld scenario observationally, we need a complete
description
of the evolution of density perturbations in the most general brane cosmological scenario
provided by this VAdS$_5$ bulk.  
With these motivations, we develop here a technique for 
relativistic density perturbations valid for this generalised brane cosmology,
which will act as a natural extension of the covariant perturbations
of General Relativistic framework \cite{GRpert} to braneworld scenario.
We further show in the subsequent discussions that the Weyl
fluid can play a crucial role in late time cosmologies as a geometric
candidate for dark matter albeit its actual material existence.


\section{Brane dynamics with Weyl fluid}

As mentioned, we shall concentrate on the most general bulk scenario,
for which 
the bulk geometry is given by a Vaidya-anti de Sitter metric
\begin{equation}
d S_5^2 = - f(r, ~v) ~dv^2 + 2 dr ~dv + r^2 d \Sigma_3^2 
\label{eqd1} 
\end{equation}
where $\Sigma_3$ is the 3-space having flat, spherical or hyperboloidal
symmetry, 
$f(r, ~v) =  k - \frac{\Lambda_5}{6}r^2 - \frac{m(v)}{r^2}$,
and $m(v)$ is the resultant of the variable mass of the Vaidya black hole and radiation field.
This type of bulk  can exchange energy with the brane as a null flow along the radial direction
\cite{maartbulk, langrev, lang1, chamb, lang2, vads5gen, asymgen, nonradial, radbh5, sustr, sucol}.
Consequently, the brane matter conservation equation is modified to 
\be
\dot\rho + 3 \frac{\dot a}{a} (\rho + p) = -2 \psi
\label{rho} 
\label{vaidden} 
\ee
where $\psi$ is the null flow characterising the VAdS$_{5}$ bulk by
 the  radiation field of a null dust $T_{AB}^{\rm bulk} = \psi q_{A} q_{B}$,
which leads to the above equation by using 
\be 
\nabla^\mu T_{\mu \nu} = -2 ~T_{AB}^{\rm{bulk}} ~n^A g_\nu^B
\ee
(where $n^A$ are the normals to the surface), which gives
\be
\nabla^\mu T_{\mu \nu} = -2 \psi u_{\mu}
\ee
($u_{\mu}$ are the unit velocity vectors), and readily leads to Eq (\ref{rho}). 
This modified conservation equation, with the help of  the Bianchi identity
on the brane $\nabla^\mu G_{\mu \nu} =0$, leads to another constraint equation
\be
\nabla^\mu {\cal E}_{\mu \nu} = \frac{6 \kappa^2}{\lambda} \nabla^\mu {\cal S}_{\mu \nu}
+ \frac{2}{3} \left[ \kappa_5^2 \left( \dot \psi + 3 \frac{\dot a}{a} \psi \right)
- 3 \kappa^2 \psi \right] u_{\mu} 
+\frac{2}{3} \kappa_5^2 \overrightarrow \nabla_\mu \psi
\ee
where $\lambda$ is the brane-tension and ${\cal E}_{\mu \nu}$ and ${\cal S}_{\mu \nu}$ are,
respectively, the projected bulk Weyl tensor and the quadratic
contribution from the brane energy-momentum tensor to the Einstein equation on the brane
\cite{sustr}.
The above equation governs the evolution of the Weyl fluid $\rho^*$ 
(so named since it is a fluid-like contribution from the bulk Weyl tensor to the brane).
For FRW geometry on the brane, this is given by
\be
\dot\rho^* + 4 \frac{\dot a}{a} \rho^* = 2 \psi - \frac{2}{3} \left ( \frac{\kappa_{5}}{\kappa} \right )^{2} 
\left( \dot \psi + 3 \frac{\dot a}{a} \psi \right)
\label{vaidweyl1} 
\ee
so that this quantity evolves as  \cite{lang2, sustr}
\be
\rho^* = \frac{C(\tau)}{a^4} \propto \frac{1}{a^{(4 - \alpha)}}
\label{vaidweyl2} 
\ee
which gives a general, physically relevant behaviour for the Weyl fluid.  
Here, $\tau$ is the proper time on the brane.
Obviously, contrary to the Sch-AdS$_5$ bulk, here 
$C(\tau)$ is evolving, and consequently, the Weyl fluid no longer 
behaves like radiation. To a brane-based observer, the cosmological dynamics
is now governed by an effective perfect fluid, the  components of which  are 
given by \cite{chamb, sustr}
\bea
\rho^{\text{eff}} &=& \rho +\frac{\rho^2}{2\lambda} + \frac{C(\tau)}{a^4} 
\label{eqb15} \\ 
p^{\text{eff}} &=& p  + \frac{\rho}{2\lambda} (\rho +2p)+\frac{C(\tau)}{3 a^4}
\label{eqb16}  
\eea
The anisotropic components of the Weyl fluid, {\em viz.}, $q^*_\mu$
and $\pi^*_{\mu\nu}$ vanish, in order that the
VAd$S_5$ bulk be compatible to FRW geometry on the brane.
The Friedmann equation and the covariant Raychaudhuri equation, expressed 
in terms of these effective quantities, are  respectively \cite{sucol}
\bea
H^2 &=& \frac{\kappa_4^2}{3} \rho^{\rm eff}
 + \frac{\Lambda}{3} - \frac{k}{a^2} 
\label{fr} \\
\dot H &=& -\frac{\kappa_4^2}{2} \left(\rho^{\rm eff} + p^{\rm eff} \right) + \frac{k}{a^2} - \frac{\kappa_5^2}{3} \psi
\label{rc} 
\eea

In the brane-based Newtonian analysis of perturbations
by considering small fluctuations of the effective density
$\rho^{\text{eff}}(\overrightarrow x, ~\tau) = \bar\rho^{\text{eff}}(\tau) 
(1 + \delta^{\text{eff}}(\overrightarrow x, ~\tau)) $
and the effective gravitational potential
$\Phi^{\text{eff}}(\overrightarrow x, ~\tau) = \Phi_0^{\text{eff}} + \phi^{\text{eff}}$ 
on the hydrodynamic equations for  this effective perfect fluid,
one obtains for a barotropic fluid a single second order equation in terms of Fourier mode 
\be
\frac{d^2 \delta_k^{\text{eff}}}{d \tau^2} + 2\frac{\dot a}{a}
\frac{d \delta_k^{\text{eff}}}{d \tau} - \left[4 \pi G \bar\rho^{\text{eff}}
- \left(\frac{c_s^{2{\text{eff}}} k}{a} \right)^2\right]\delta_k^{\text{eff}} = 0
\label{eqc11} 
\ee
where $c_s^{2{\text{eff}}}$ is  the square of the effective sound speed \cite{maartrev, sustr}.
The above perturbation equation of the effective fluid can 
account for the required amount of gravitational instability  if the Weyl density 
 redshifts more slowly than baryonic matter density, so that  it can 
dominate over  matter at late times, which is realised when $1 < \alpha < 4$
in Eq (\ref{vaidweyl2}). 
Now, for late time behaviour, we can drop the quadratic terms in equations (\ref{eqb15}) and (\ref{eqb16})
so that the effective density is given by 
$\rho^{\text{eff}} = \rho^{(b)} + \rho^*$ 
which is now constituted of the usual matter (baryonic) density $\rho^{(b)}$ 
and an additional density contribution from the Weyl fluid.
This Weyl density, being geometric, is essentially non-baryonic. Consequently,
we can decompose Eq (\ref{eqc11}) to get the individual evolution
equations for the perturbation for each of the fluids 
\begin{eqnarray}
\frac{d^2 \delta^{(b)}}{d \tau^2} + 2\frac{\dot a}{a} \frac{d \delta^{(b)}}{d \tau} 
= 4 \pi G \bar\rho^{(b)} \delta^{(b)} + 4 \pi G \bar\rho^* \delta^*
\label{eqc18} \\
\frac{d^2 \delta^*}{d \tau^2} + 2\frac{\dot a}{a} \frac{d \delta^*}{d \tau} 
= 4 \pi G \bar\rho^* \delta^* + 4 \pi G \bar\rho^{(b)}  \delta^{(b)} 
\label{eqc19} 
\end{eqnarray}
where $\delta^{(b)}$ and $\delta^*$ are the fluctuations of baryonic
matter and Weyl fluid respectively. 
With $\Omega^{(b)}  \ll \Omega^*$, the relevant growing mode solutions are given by \cite{sustr}
\bea
\delta^*(z) &=& \delta^*(0) (1 + z)^{-1}
\label{eqc20} \\
\delta^{(b)}(z) &=& \delta^*(z) \left(1 - \frac{1+z}{1+z_N}\right) 
\label{eqc23} 
\eea
with the input that the late time behaviour of the expansion of the universe
in RS II is the same as the standard cosmological solution for the 
scale factor \cite{maartbulk, maart6}
where  the scale factor is related to the redshift function by $a \propto (1+z)^{-1}$.

The solutions reveal that at a redshift close to $ z_N$, 
the baryonic fluctuation $\delta^{(b)}$ almost vanishes but the Weyl fluctuation 
$\delta^*$ still remains finite. So, even if the baryonic fluctuation 
is very small at a redshift of $z_N \approx 1000$,
as confirmed by CMB data \cite{cmb}, the fluctuations of the Weyl fluid still
had a finite amplitude during that time, whereas
at a redshift much less than  $z_N$
the baryonic matter fluctuations are of equal amplitude as
the Weyl fluid fluctuations. This is precisely what is required to
explain the formation of structures we see today.
Thus, the Newtonian analysis of perturbations on the brane is capable of
explaining structure formation (within its limit) by Weyl fluid,
devoid of any material existence of dark matter. Hence, the Weyl fluid
acts as a possible geometric candidate for dark matter.


\section{Relativistic  perturbations with Weyl fluid}

The Newtonian analysis depicted so far turn out to be an useful
tool to study perturbations on the brane after the  universe enters the Hubble length.
A more complete picture can be obtained
if one studies relativistic analysis of perturbations, which include the
evolution of the universe at the  super-Hubble scale as well.
In this section we shall develop a multi-fluid perturbative technique
in order to discuss relativistic perturbation relevant for brane cosmology. 
This will be carried through in the subsequent 
sections for the purpose of analysis for different braneworld scenarios.
Our basic motivation in the attempt to develop a multi-fluid
perturbative technique is governed by the realisation obtained from Newtonian analysis that,
contrary to the Sch-AdS$_5$ bulk scenario,
the Weyl fluid may not be that much insignificant
so as to neglect its effects at late time, if we have a general  bulk geometry.
Consequently, in a general  brane cosmological scenario, along with baryonic matter,
 the universe consists of a  considerable amount of Weyl fluid as well.

Before going into the details,
let us jot down here the major points in addressing relativistic perturbations on the brane.
\begin{itemize}
\item Here the cosmological dynamics is governed by a  two-fluid system.
One of the components of the system is a material fluid  $\rho^{(b)}$ -- the baryonic matter content on the brane. The second component is a
 geometric fluid $\rho^{*}$ --  the Weyl fluid.
The total (effective) density for the system on the brane is given by 
$\rho^{\rm eff} = \rho^{(b)} + \rho^{*}$.

\item Though there are two components of the effective fluid,
the Weyl fluid being a geometric entity, there is  a single material fluid
in the analysis. As a results, there will be no entropy perturbation as such.

\item For the same reason, there is no peculiar velocity for the Weyl component,
leading to $v^* =0$. 

\item The anisotropic components of the Weyl fluid being absent so as to fit
it into an FRW background, we will set $q_{\mu}^* = 0 = \pi_{\mu\nu}^*$ right
from the beginning. As a result, each component
of the two-fluid system behaves individually like a perfect fluid, resulting
in a perfect fluid behaviour for the effective fluid as a whole.

\item These two fluids interact and exchange energy between them,
which is governed via the bulk-brane energy exchange 
and the backreaction of the system on the brane.

\item Since there is energy exchange between these two fluids, the conservation
equation for each individual is now modified.
These modified forms of the conservation equations have been explained in 
equations (\ref{vaidden}) and (\ref{vaidweyl1}).
\end{itemize}

Because of the interaction between the two fluid components, each of the  
two modified conservation equations 
 can be  written  in terms of the contribution from the interaction as
\begin{equation}
\dot\rho^{(i)} + \Theta (\rho^{(i)} + p^{(i)}) = I^{(i)}
\label{int} 
\end{equation}
where $\Theta = 3\frac{\dot a}{a}$ is the volume expansion rate,
a superscript $(i)$ denotes the quantities for the $i$-th fluid and $I^{(i)}$ is the corresponding interaction term. It readily follows from equations (\ref{vaidden}) and (\ref{vaidweyl1}) that 
the interaction terms, when written explicitly, are given by
\begin{eqnarray}
I^{(b)} &=& - 2 \psi \\
I^{*} &=& 2 \psi - \frac{2}{3} \left ( \frac{\kappa_{5}}{\kappa} \right )^{2} 
\left( \dot \psi + 3 \frac{\dot a}{a} \psi \right) 
\end{eqnarray}

For relativistic perturbations,  we express the densities of each
of the contributing fluid components in terms of dimensionless parameters as
\be
\Omega_{\rho ^{(b)}} =\frac{\kappa ^{2}\rho^{(b)}}{3H_{0}^{2}} ~~ , ~~
 \Omega_{\rho^{*}} =\frac{\kappa ^{2}\rho^*}{3H_{0}^{2}}
\label{param} 
\ee
with the first one  for baryonic matter and the second one for Weyl fluid.
Considering the nature of the Weyl fluid as discussed in the previous section, we find that
the density parameter for the Weyl fluid is given by 
\be
\Omega_{\rho^{*}}=\frac{2C_{0}}{a_{0}^{4-\alpha}H_{0}^{2}}
\ee
where $C_{0}$ is the onset value for the Weyl parameter $C(\tau)$.

For completion, we mention here that there can, in principle, appear
 two more dimensionless parameters,
one each for the cosmological constant and the brane tension arising in the brane cosmological context.
They are
\be
 \Omega _{\Lambda }=\frac{\Lambda }{3H_{0}^{2}} ~~ , ~~
\Omega _{\lambda} =\frac{\kappa ^{2}\rho _{0}^{2}}{6\lambda H_{0}^{2}}
\label{paramir} 
\ee
with the total density satisfying the critical value
\be
\Omega _{\rm tot} =\sum_{i} \Omega_i 
= \Omega_{\rho ^{(b)}} +  \Omega_{\rho^{*}} +  \Omega _{\Lambda } + \Omega _{\lambda}
=1
\ee
Here, and throughout the rest of this article, we have considered a spatially flat
universe with $k=0$.
In Eq (\ref{paramir}), the first one is relevant if one considers cosmological constant
in this brane universe while studying the expansion history of the universe
whereas the one due to the brane tension is relevant in the high energy
early universe (inflationary) 
phase but is negligible for low energy late time phenomena such as structure formation.
Thus the baryonic density and the Weyl density are the only two relevant contributions
in the scenario being discussed here.
In what follows we shall restrict ourselves to the discussion of 
the  Einstein-de Sitter brane universe for which $\Omega_\Lambda = 0$
leading to $\Omega _{\rm tot} \approx \Omega_{\rho ^{(b)}} +  \Omega_{\rho^{*}} = 1$.

We now express  the comoving fractional gradients of the effective density and expansion
relevant in the brane cosmology as 
\begin{eqnarray}
\Delta^{(i)}_{\mu} &=& \frac{a}{\rho^{(i)}} D_{\mu} \rho^{(i)}\\
Z_{\mu}^{\rm eff}
 &=& a D_{\mu} \Theta \\
\Delta_{\mu}^{\rm eff} &=& \frac{a}{\rho^{\rm eff}} D_{\mu} \rho^{\rm eff}
\end{eqnarray}

As already discussed, both baryonic matter and Weyl fluid behave individually as perfect
fluid components, which means the effective flux arising from the peculiar velocities
of each component vanish to zero order, confirming that the perturbations considered
here are gauge-invariant at the first order.

With the above notations, the linearised evolution equation for the density
perturbations in the braneworld is obtained by taking spatial gradient
of the modified conservation equations. After linearisation, it turns out to be
\be
\dot\Delta^{(i)}_{\mu}   =  \left ( 3 H w^{(i)} -  \frac{I^{(i)}}{\rho^{(i)}}     \right ) \Delta_{\mu}^{(i)} - (1 + w^{(i)}) Z_{\mu}^{\rm eff}  
 -  \frac{c^{2{\rm eff}}_{s} ~ I^{(i)} }{\rho^{(i)} (1 + w^{\rm eff})}  \Delta_{\mu}^{\rm eff} - \frac{3 a H I^{(i)}_{\mu}}{\rho^{(i)}}  + \frac{a}{\rho^{(i)}} D_{\mu} I^{(i)} 
\label{dengrad}  
\ee
where $w^{(i)} = p^{(i)}/ \rho^{(i)}$ is the equation of state for $i$-th fluid
and $c_{s}^{2(i)} = \dot p^{(i)}/ \dot\rho^{(i)}$ is the sound speed squared for that species,
with the corresponding quantities for the effective total fluid are, respectively,
\bea
w^{\rm eff} &=& \frac{1}{\rho^{\rm eff}} \sum_{i} \rho^{(i)} w^{(i)}  \\
c_{s}^{2{\rm eff}} &=& \frac{1}{\rho^{\rm eff}(1+w^{\rm eff})} \sum_{i} c_{s}^{2(i)} \rho^{(i)} (1+ w^{(i)})
\eea

In the relativistic perturbations, contrary to the Newtonian analysis, we further
have an evolution equation for the effective expansion gradient, which depends on 
the effective fluid. This is obtained by taking spatial gradient of the modified
Raychaudhuri equation (\ref{rc}) 
and is given in the braneworld scenario by
 \be
\dot Z_{\mu}^{\rm eff}  +  2 H Z_{\mu}^{\rm eff}  =  -  \frac{\kappa^{2}}{2} \rho^{\rm eff} \Delta^{\rm eff} - \frac{c^{2{\rm eff}}_{s}}{1 + w^{\rm eff}} D_{\mu} D^{\nu} \Delta_{\nu}^{\rm eff}  
 + \frac{\kappa^{2}_{_{5}} \psi}{1 + w^{\rm eff}}  c^{2{\rm eff}}_{s} \Delta_{\mu}^{\rm eff} - a \kappa^{2}_{_{5}} D_{\mu} \psi
\label{expgrad} 
\ee

It should be mentioned here that since the evolution of the expansion gradient
is dependent on the curvature perturbations, the later should not remain strictly constant
in this multi-fluid perturbation scenario. However, though there is a significant energy 
exchange between  brane matter and  Weyl fluid at early times, we shall see from the next 
section that the energy exchange between the two fluids are almost in equilibrium 
at late times, so that the local curvature perturbations can safely be considered 
to be constant for all practical purpose. One should, however, consider 
the variation of this term while analysing inflationary phase for an instance.
We follow this argument right from here
in order to avoid mathematical complexity.

As in GR, we find that while discussing perturbations in brane cosmology,
it is advantageous to express 
the above equations in terms of covariant quantities. 
These  density perturbations are governed by the 
fluctuation of the following covariant projections 
\begin{eqnarray}
\Delta^{(i)} &=& a~ D^{\mu} \Delta^{(i)}_{\mu} \\
 \Delta^{\rm eff} &=& a~ D^{\mu} \Delta_{\mu}^{\rm eff} \\
  Z^{\rm eff} &=& a~ D^{\mu}Z_{\mu}^{\rm eff}
\end{eqnarray}

Consequently, the covariant density perturbation equation and
expansion gradient on the brane, when expressed in terms of
the above covariant quantities, are obtained straightaway from equations (\ref{dengrad}) 
and (\ref{expgrad}). They are given by 
\begin{eqnarray}
\dot\Delta^{(i)} & = & \left ( 3 H w^{(i)} - \frac{I^{(i)}}{\rho_{(i)}} \right ) \Delta^{(i)} - (1 + w^{(i)}) Z^{\rm eff}  
- \frac{c^{2{\rm eff}}_{s} ~ I^{(i)} }{\rho^{(i)} (1 + w^{\rm eff})}  \Delta^{\rm eff}
\nonumber \\
& - & \frac{3 a^{2} H D^{\mu} I^{(i)}_{\mu}}{\rho^{(i)}}  + \frac{a^{2}}{\rho^{(i)}} D^{2} I^{(i)} 
\label{covdengrad} \\
 \dot Z^{\rm eff} & + & 2HZ^{\rm eff} = - \frac{\kappa^{2}}{2} \rho^{\rm eff} \Delta^{\rm eff} - \frac{a c^{2{\rm eff}}_{s}}{1 + w^{\rm eff}} D^{2} \Delta^{\rm eff} 
+ \frac{\kappa^{2}_{_{5}} \psi}{1 + w^{\rm eff}}  c^{2{\rm eff}}_{s} \Delta^{\rm eff} - a^{2} \kappa^{2}_{_{5}} D^{2} \psi
\label{covexpgrad} 
\end{eqnarray}
In deriving the above covariant perturbation equations, we have considered those kind
of perturbations for which,
like the  unperturbed Weyl fluid, 
the anisotropic stresses and fluxes for the perturbed Weyl fluid are also vanishing.  

These set of equations provide the key information about the perturbation in brane cosmology.
In the subsequent section, we shall try to analyse these relativistic perturbation
equations and obtain possible consequences.


\section{Solutions and Analysis}


\subsection{Empty bulk : Non-interacting fluids}

Let us now discuss the special scenario when the bulk is empty for which
the VAdS$_5$ bulk reduces to Sch-AdS$_5$.
In this case, there is no question of energy exchange
 between the brane and the 
bulk. Consequently, there is no interaction between brane matter and Weyl fluid as such,
which reveals from Eq (\ref{int}) the fact that the individual conservation equation 
for each of the components are preserved. Thus, the Weyl fluid evolves in this case as
\be
\rho^* \propto a^{-4}
\label{darkrad} 
\ee
with the Weyl parameter $\alpha$ now being zero, so that for empty bulk,
 the Weyl fluid behaves like radiation, for which this is called dark radiation. 

Since in this case, there is no interaction between the
two components of the effective fluid and also, there is no null flow from  the bulk 
to the brane (or vice versa), 
we can drop the interaction terms and the terms involving  $\psi$ 
in the analysis. As a result, the covariant perturbation equations (\ref{covdengrad})
and (\ref{covexpgrad}) are vastly simplified. They are now given by
\begin{eqnarray}
\dot\Delta^{(i)} & = & 3 H w^{(i)} \Delta^{(i)} - (1 + w^{(i)}) Z^{\rm eff} \\ 
 \dot Z^{\rm eff} & + & 2HZ^{\rm eff} = - \frac{\kappa^{2}}{2} \rho^{\rm eff} \Delta^{\rm eff} - \frac{a c^{2{\rm eff}}_{s}}{1 + w^{\rm eff}} D^{2} \Delta^{\rm eff} 
\end{eqnarray}

Taking the time derivative of the above two equations and combining them,
we  obtain the evolution equations for density perturbations of the two  fluids
\bea
\ddot\Delta^{(b)} + 2 H \dot\Delta^{(b)} &=& \frac{\kappa^{2}}{2} \rho^{\rm eff} \Delta^{\rm eff}  \\
\ddot\Delta^{*} + 2 H \dot\Delta^{*} &=& \frac{4}{3} \frac{\kappa^{2}}{2} \rho^{\rm eff} \Delta^{\rm eff}  
+ \Delta^{*} \left( 2H^{2} - \frac{\kappa^{2} }{2} \right) 
 + H \dot\Delta^{*} 
\eea
where the first equation is for baryonic matter while the second one for dark radiation.

Recall that the amount of dark radiation is constrained by the Nucleosynthesis data
to be at most $3\%$ of the total radiation density of the universe.
So, it redshifts at a faster rate than ordinary matter on the brane
so that the matter on the brane becomes 
dominant on the Weyl fluid at late time.
Hence, it is expected that the dark radiation  does not play any significant role in late
time cosmologies. 
It is obvious from the fact that in this case, $\Omega^{(b)} \gg \Omega^*$,
which when put back into the above equations, leads to
$\Delta^{(b)} \gg \Delta^*$,
so that the dark radiation fluctuation does not contribute substantially
at late times.
The Sch-AdS$_5$ bulk scenario thus fails to explain structure formation with only baryonic matter
and dark radiation. One needs cold dark matter in the theory
and  the dark radiation
can, at best, slightly modify the standard perturbative analysis.

\subsection{General bulk : Interacting Weyl fluid}

The general scenario, however, is different from the empty bulk case
since now the Weyl fluid  exchanges energy with brane matter through
interactions and is the dominant contribution of the effective fluid
in the perturbation equations.
Here, the evolution of perturbations for the individual fluids  are governed by equations (\ref{covdengrad}) 
and (\ref{covexpgrad}), which now include the effects of the interaction terms 
as well as of the effect of null radiation through the term involving $\psi$.
With these inclusions, 
the equations become a bit
 too complicated and it is almost impossible to have an analytical solution
from these complicated equations. However, the equations turn out to be tractable 
if we incorporate certain simplifications following physical arguments,
without losing any essential information as such.
The simplifications we incorporate are as follows: 
\begin{itemize}
\item The null flow from the brane to the bulk $\psi$ 
is a function of time only. This means that we
are considering only the time-evolution for the null radiation, at least on the brane,
 which is relevant for its late time behaviour in perturbation analysis. 

\item The energy exchange between the two fluids is in equilibrium, i.e., 
the energy received by the Weyl fluid is the same as 
the energy released by brane matter, so that $\sum_{i} I^{(i)} = I^{(b)} +I^{*}=0$.
Hence, no extra energy is
leaked to the bulk from the brane at late time (though at early time there may be some
leakage of energy from the brane to the bulk). 
 This basically describes the late time behaviour, consistent with the fact
that the standard evolution history (scale factor) are regained in this scenario 
at the ``matter-dominated'' era \cite{maartbulk, maart6}.
\end{itemize}

Now, we have shown that in this generalised braneworld scenario,
the Weyl fluid, in general, evolves as (Ref Eq (\ref{vaidweyl2}))
\begin{equation}
\rho^* = C_0 a^{-(4 - \alpha)}
\end{equation}
with the parameter $\alpha$ in the range $1 < \alpha <4$ 
so that it is the dominant contribution in the two-fluid system.
The energy exchange between the components of the system being in equilibrium,
we find from Eq (\ref{vaidweyl2}) that
the Weyl fluid now behaves as
\be
\rho^{*} \propto {a^{-3/2}}
\label{weyl3/2} 
\ee
with the parameter $\alpha = \frac{5}{2}$.
This readily suggests that the Weyl fluid actually redshifts more slowly than ordinary matter
and hence, can dominate over matter at late times, reflecting one of the fundamental 
properties of dark matter.
This also provides a more stringent bound for the value of $\alpha$ from theoretical ground alone
(which was predicted from Newtonian analysis to fall within 1 to 4).

We now take the time derivative of the covariant perturbation equations
 (\ref{covdengrad}) and (\ref{covexpgrad}), and rearrange terms so as to obtain
 a single second order differential equation 
for each of the fluids.
Thus, the 
equation describing evolution of scalar perturbations of matter on the brane turn out to be
\be
\ddot\Delta^{(b)} + 2 H \dot\Delta^{(b)} = \frac{\kappa^{2}}{2} \rho^{\rm eff} \Delta^{\rm eff}  - \frac{c^{2{\rm eff}}_{s} \kappa^{2}_{5} \psi}{1 + w^{\rm eff}}   \Delta^{\rm eff} 
+ \frac{4H \psi}{\rho^{(b)}} \left( \Delta^{(b)} + \frac{ c^{2{\rm eff}}_{s} \Delta^{\rm eff}}{1 + w^{\rm eff}}  \right) 
+ \left[ \frac{2 \psi}{\rho^{(b)}}  \left( \Delta^{(b)} + \frac{ c^{2{\rm eff}}_{s} \Delta^{\rm eff}}{1 + w^{\rm eff}}  \right) \right] ^{\cdot} 
\label{barypert} 
\ee
whereas the scalar perturbation equation for the  Weyl fluid on the brane is given by
\bea
\ddot\Delta^{*} &+& 2 H \dot\Delta^{*} = \frac{4}{3} \frac{\kappa^{2}}{2} \rho^{\rm eff} \Delta^{\rm eff}  
 - \frac{c^{2{\rm eff}}_{s} \Delta^{\rm eff}}{1 + w^{\rm eff}}  \left( \frac{7H\psi}{\rho^{*}} + 
\frac{4  \kappa^{2}_{_{5}} \psi }{3} + \frac{2 \dot\psi}{\rho^{*}} \right) \nonumber \\ 
&-& \frac{c^{2{\rm eff}}_{s} \dot\Delta^{\rm eff}}{1 + w^{\rm eff}} \frac{2 \psi}{\rho^{*}}
+ ~\Delta^{*} \left( 2H^{2} - \frac{\kappa^{2} }{2} 
- \frac{7H \psi}{\rho^{*}} - \frac{2 \dot\psi}{\rho^{*}} \right) 
 + \dot\Delta^{*} \left(H - \frac{2 \psi}{\rho^{*}} - \frac{\kappa^{2}_{_{5}} 
 \psi}{3}\right)
\label{weylpert} 
\eea

Recall from the discussions following Eq (\ref{weyl3/2}) 
that in this scenario the Weyl fluid is the dominant component
of the effective fluid. Consequently, 
the evolution equation for the Weyl fluid at late times is radically
simplified by using $\Delta^{(b)} << \Delta^{*}$ since the Weyl fluid is now the dominant contribution.
With the energy exchange between the two fluids being in equilibrium,
the expression for the null flow further simplifies the above equation
so that it can  now be recast
in the following form 
\begin{equation}
 \ddot\Delta^{*} + \frac{A}{t} \dot\Delta^{*} - \left( \frac{{B}}{t} +\frac{{C}}{t^2} \right) \Delta^{*} = 0
\label{final} 
\end{equation}
where the constants $A, B, C$ are readily determined from the constraint equations.
These constants are given by
\bea
A &=& \frac{2}{3} + \frac{5}{2} \left( \frac{\psi_0}{\rho_0^*}\right) 
\left(\frac{2}{3} \frac{a_0}{H_0}\right)^{2/3}\\
B &=&  \frac{2}{3} \kappa^2 \rho_0 \left(\frac{2}{3} \frac{a_0}{H_0}\right)^{3/2}
+ \left(1+ \frac{\kappa^2}{6} \rho_0^* \right) 
\left(\frac{2}{3} \frac{a_0}{H_0}\right)^{2/3} \\
C &=& \frac{A}{4} - \frac{19}{18}
\eea

The above equation (\ref{final}) for $\Delta^{*}$ turns out to be somewhat tractable. 
One of its solutions is given by 
\begin{equation} 
 \Delta^{*} \sim t^{\frac{1}{2}-\frac{A}{2}} {\rm Bessel}I \left[ \sqrt{1 - 2A + A^2 + 4C}, 2 \sqrt{B} \sqrt{t} \right] 
 \end{equation}
The above solution, consisting of a Bessel function, is found to be a growing function.
Therefore, the evolution equation for the Weyl fluid, indeed, shows a growing mode solution,
which is required to explain the growth of perturbations at late times.
Thus, the relativistic perturbation theory relevant in brane cosmology
 gives rise to a fluid which is very
different from dark matter in origin and nature but has the potentiality
to play the role of dark matter in cosmological context. It is worthwhile to note
that, to a brane-based observer, the nature of the Weyl fluid is determined from
bulk geometry  arising from the radiation flow in the bulk.
That is why the Weyl fluid can be treated as a geometric candidate for dark matter.

The following figure depicts a qualitative behaviour of the growth of Weyl fluid perturbations
with time. 
\begin{figure}[htb]
{\centerline{\includegraphics[width=7cm, height= 4.5cm] {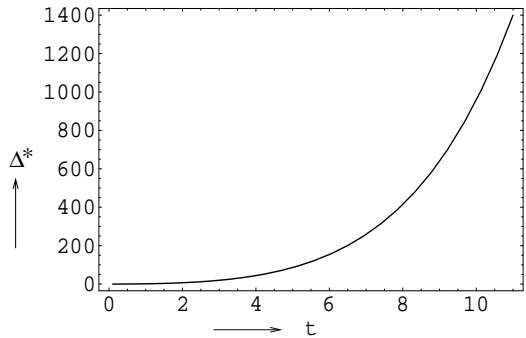}}}
\caption{Growth of Weyl fluid perturbations with time}
\end{figure}
The figure once again shows that the evolution of perturbations of Weyl fluid
is very different from cold dark matter (CDM), which makes the theory distinct from
standard analysis with CDM. We, however, note that
since the dynamics here is completely different from the standard one involving CDM, 
one cannot comment conclusively on the merits/demerits of 
Weyl fluid over CDM  right from here.
One has to  reformulate and estimate different cosmological parameters in this context and
confront them with observations for a more conclusive remark.
For example, the relation of the transfer function with the potential 
will now be replaced by a novel relation  
with the effective potential discussed in the brane cosmological context
\cite{sustr}. As a result, the variation of the growth function with the
scale factor may not be the same as usually needed in the standard cosmological paradigm.
It is to be seen if this analysis of perturbation with the Weyl fluid fits
in this new, brane cosmological framework, which is not a trivial exercise, we suppose. 
The interested reader may further refer to \cite{dodelson} for 
an overall view on how different cosmological parameters are developed
in a specific theoretical framework.

However, even at this stage, our model does show some agreement with observational results.
From the recent studies on confronting braneworld models with observations \cite{obs}
by obtaining the luminosity distance for FRW branes with the Weyl fluid,
it is found that a certain amount of
Weyl fluid with $2 \leq \alpha \leq 3$ is in nice agreement with Supernovae data.
From the relativistic perturbations discussed in this article we have found a specific value for
$\alpha$, namely
 $\alpha = \frac{5}{2}$, which falls within this region.
Thus the braneworld model of perturbations fits well in this observational scenario. 
We hope  an extensive study in this direction will lead to more interesting results
to make a more conclusive remark.


\section{Summary and open issues}

In this article, we have developed a technique for relativistic  perturbations
valid for a general brane cosmological scenario. The essential distinction of our analysis
from the studies on brane cosmological perturbations 
available in the literature is that, here the geometrical effect of the bulk on the brane --
the so-called Weyl fluid  -- plays a very crucial role in determining the nature of
the evolution of density perturbations. This is materialised from the realisation that
in the general brane cosmological scenario obtained from Vaidya-anti de Sitter bulk,
the Weyl fluid plays a significant role in controlling the dynamics on the brane,
contrary to the earlier results based on dark radiation. Our results are, in a sense,
a generalisation of the multi-fluid covariant perturbation formalism in brane cosmological
framework. Further, we have solved the perturbation equations and found that  
the perturbation of the Weyl fluid grows at late time, and thus, this component
of braneworld gravity plays a significant
 role in late time cosmology to act as a possible geometric candidate for dark matter.
We have discussed some of the implications of fluctuations involving it and  
have   mentioned some observable sides of this model as well.

An important issue is to fit this theoretical model with current
observational data. 
Recently there has been some progress in this direction \cite{obs}.
An extensive study 
on confronting this braneworld model with observations in a more rigorous method 
can provide us with necessary information on the merits and demerits of the formalism.
To this end, a thorough study of different parameters related to cosmological perturbation
is to be performed. 
As mentioned in the previous section, the different cosmological parameters need
to be reformulated in this framework. The next step is to estimate them 
and confront them with observations. For example, it is to be seen if the power spectrum,
redefined in this paradigm 
with the Weyl fluid acting as a dark matter candidate, fits with 
the highly accurate observational data.

Further, analysis of different types of metric-based perturbations, namely, scalar,
vector and tensor as well as related issues like CMB anisotropy, Sachs-Wolfe effect
etc has to be be studied in details in this brane cosmological framework
 with a significant  Weyl fluid.
An extensive study in this direction is essential,
which we hope to address in near future.
Also, to apply this formalism in the framework of branworld models of 
dark energy  \cite{de} 
remains as  another interesting issue.


\section*{Acknowledgements} 

The author  thanks Biswajit Pandey, Ratna Koley and Varun Sahni for 
stimulating discussions.


\end{document}